\documentclass[preprint]{elsarticle}
\usepackage{times}
\usepackage{graphicx}
\usepackage{bm}
\usepackage{multirow}
\usepackage{subfigure}
\usepackage{amsmath,amssymb}

\newcommand{\abs}[1]{\left\lvert #1\right\rvert}

\newcommand{\sign}{\text{sgn}}
\newcommand{\up}{\uparrow}
\newcommand{\down}{\downarrow}

\newcounter{lastnote}

\def\bc{\begin{center}}
\def\ec{\end{center}}
\def\be{\begin{equation}}
\def\ee{\end{equation}}

\begin{document}
\title{The stability of the fractional quantum Hall effect in topological insulators}
\author[cns]{Ashley M. DaSilva\corref{cor1}}
\ead{amd405@psu.edu}
\cortext[cor1]{Corresponding author}
\address[cns]{The Center for Nanoscale Science and Department of Physics, The Pennsylvania State University, University Park, Pennsylvania 16802-6300, USA}
\begin{abstract}
With the recent observation of graphene-like Landau levels at the surface of topological insulators, the possibility of fractional quantum Hall effect, which is a fundamental signature of strong correlations, has become of interest. Some experiments have reported intra-Landau level structure that is suggestive of fractional quantum Hall effect. This paper discusses the feasibility of fractional quantum Hall effect from a theoretical perspective, and argues that while this effect should occur, ideally, in the $n=0$ and $|n|=1$ Landau levels, it is ruled out in higher $|n|$ Landau levels. Unlike graphene, the fractional quantum Hall effect in topological insulators is predicted to show an interesting asymmetry between $n=1$ and $n=-1$ Landau levels due to spin-orbit coupling.
\end{abstract}

\maketitle

\section{Introduction}
The fractional quantum Hall effect (FQHE) is a fundamental signature of strong correlations in two dimensional systems. In the FQHE, the electrons lower their interaction by binding with vortices to form composite fermions~\cite{Jain1989,JainText} which are then weakly interacting. The integer quantum Hall effect of these composite fermions is the FQHE of the electrons.~\cite{Jain1989,JainText} The formation of composite fermions and the FQHE relies on an ineraction between electrons that is sufficiently short range. The composite fermion wavefunction is a good approximation when this interaction is sufficiently short ranged, a condition which can be quantified by the Haldane pseudopotentials.~\cite{JainText,Quinn2000,Quinn2000b,Peterson2008}

With rapid improvement in the quality of topological insulators (TIs) has come progress towards the observation of integer and fractional quantum Hall effects in these systems. Topological insulators have insulating bulk electronic states but an odd number of topologically protected Dirac cones on the surface.~\cite{Fu2007, Moore2007, Roy2009} Because of the Dirac nature of the electrons, many properties of TIs are analagous to those of graphene. One of these properties is the Landau level (LL) structure for Dirac fermions, which is quite unique; the energy levels are proportional to the square root of the LL indices due to the chiral nature of the electrons.~\cite{McClure1956, Novoselov2005, Zhang2005, CastroNeto2009, Yang2011} Recently this graphene-like integer quantum Hall effect has been observed in strained bulk HgTe~\cite{Brune2011} and the LL spectrum has been measured in Bi$_{2}$Se$_{3}$ using scanning tunneling microscopy~\cite{Cheng2010,Hanaguri2010}. In addition, features in the Hall resistance at fractional filling factors have been speculated to be related to the FQHE of topological insulators.~\cite{Analytis2010,Xiong2011} With these experiments, coupled with the observation of the FQHE in graphene~\cite{Du2009,Bolotin2009} and the evidence for a strong inter-electron interaction in TIs~\cite{Wang2011}, arises the question of the possiblity of observing the FQHE in topological insulators.

This paper reports on the stability of the FQHE in topological insulators from a theoretical perspective. It is shown that the FQHE is observable for $n=0$ and $\abs{n}=1$ LLs, however is not stable for higher LLs. Due to the spin orbit interaction of the TIs, there is an anisotropy of the $\abs{n}=1$ LLs which is not seen in graphene or in the lowest ($n=0$) Landau level.~\cite{Toke2006,Apalkov2006,Goerbig2006,Nomura2006} These results hold when the bulk is conducting, although a bulk contribution to the conductivity may make these features more difficult to observe.

\section{Theory}

The Landau levels of topological insulators have been calculated~\cite{Kane2007,Mondal2010} and bear a striking similarity to those of graphene~\cite{Toke2006,McClure1956,CastroNeto2009,Apalkov2006}. There are however two important differences. First, topological insulators have a high Zeeman energy, with a $g$-factor as large as 30 (Ref.~\cite{Kohler1975}). Second, although the Hamiltonians of a TI and graphene are mathematically identical, the Hilbert space basis functions are different. In graphene, spin up and spin down states are degenerate, and the Hamiltonian is written in the pseudospin basis, which is related to the presence of two atoms in the unit cell of the hexagonal lattice. In TIs, the Hamiltonian is written in the basis of the real spin. These two factors manifest as a tendency for the spin to polarize along the direction of magnetic field in TIs. This work considers the ideal case of a single Dirac cone, occuring in materials such as Bi$_2$Se$_3$ and Bi$_2$Te$_3$,~\cite{Hsieh2009,Zhang2009} and restrict the calculation to a single surface. If the two surfaces are not too close together, the final results will hold for each surface independently.

The stability of the FQHE states can be predicted by considering the ratio of the Haldane pseudopotentials, which are the matrix elements of the interaction of the electrons within a LL,~\cite{JainText,Quinn2000,Quinn2000b,Peterson2008}
\begin{equation}
V_{\rm eff}^{nm}=\langle\Psi_{nm}\lvert V\rvert \Psi_{n0}\rangle
\end{equation}
where $\lvert\Psi_{nm}\rangle$ is the state in the $n$th LL with angular momentum $-\hbar m$ and $V$ is the interaction. These pseudopotentials measure the interaction between particles in two states within the same LL. The two-body states must be antisymmetric which implies that only pseudopotentials of odd relative angular momentum, ($-\hbar m$ where $m$ is odd) are relevant. 
Typically one can predict the stability of the FQHE by considering only the $m=1$ and $m=3$ pseudopotentials, since $V_{\rm eff}^{m}$ falls off quickly as $m$ increases. Although there are ways to more precisely quantify the pseudopotentials' role in predicting the stability of the composite fermions, these states tend to be stable if the ratio of the $m=1$ pseudopotential to the $m=3$ pseudopotential is larger than $1.3-1.5$ (Ref.~\cite{JainText,Quinn2000,Quinn2000b,Peterson2008}); in this case the FQHE may be observed. This is because the composite fermion wavefunction is accurate when the interaction is short range ($V_{\rm eff}^{1}\neq 0$, $V_{\rm eff}^{m>1}=0$), and remains a good approximation when $V_{\rm eff}^{1}/V_{\rm eff}^{3}$ is large enough.

The non-interacting electron Hamiltonian in the presence of a perpendicular magnetic field is~\cite{Mondal2010}
\begin{equation}
H=v_{F}{\bm \sigma}\cdot {\bm \Pi}+g\mu_{B}B\sigma_{z}
\end{equation}
where $v_{F}$ is the Fermi velocity and ${\bm \Pi}={\bm p}+e{\bm A}/c$ is the two-dimensional canonical momentum. The $z$-direction is taken to be perpendicular to the surface and the symmetric gauge, ${\bm A}=(B/2)(-y, x,0)^{T}$, has been used. Here, the superscript $T$ stands for transpose. In analogy with graphene~\cite{Toke2006,CastroNeto2009} this Hamiltonian can be written in terms of ladder operators,
\begin{equation}
H=\left(\begin{array}{cc}
g\mu_{B}B & -i\hbar v_{F}\ell_{B}^{-1}\sqrt{2} \: a\\
i\hbar v_{F}\ell_{B}^{-1}\sqrt{2}\: a^{\dagger} & -g\mu_{B}B
\end{array}\right)
\end{equation}
where $\ell_{B}=\sqrt{eB/\hbar c}$ is the magnetic length and
\begin{equation}
\begin{split}
a^{\dagger}=&\frac{1}{\sqrt{2}}\left(\frac{\bar{z}}{2\ell_{B}}-2\ell_{B}\partial_{z}\right)\\
a=&\frac{1}{\sqrt{2}}\left(\frac{z}{2\ell_{B}}+2\ell_{B}\partial_{\bar{z}}\right)\\
\end{split}
\end{equation}
are the ladder operators. Here, $z=x+iy$, as usual. There are two more linearly independent ladder operators,~\cite{JainText}
\begin{equation}
\begin{split}
b^{\dagger}=&\frac{1}{\sqrt{2}}\left(\frac{z}{2\ell_{B}}-2\ell_{B}\partial_{\bar{z}}\right)\\
b=&\frac{1}{\sqrt{2}}\left(\frac{\bar{z}}{2\ell_{B}}+2\ell_{B}\partial_{z}\right)
\end{split}
\end{equation}
which are related to the $z$-component of the angular momentum, $L_{z}=-\hbar(b^{\dagger}b-a^{\dagger}a)$. In analogy with graphene, the solutions are linear combinations of the usual two dimensional electron gas (2DEG) eigenstates,~\cite{JainText} $\phi_{nm}$, which satisfy $a^{\dagger}a\phi_{nm}=n\phi_{nm}$ and $b^{\dagger}b\phi_{nm}=m\phi_{nm}$,
\begin{equation}
\Psi_{n,m}(r)= \left(\begin{array}{c}
\alpha_{n}\phi_{\abs{n}-1,m}\\
\beta_{n}\phi_{\abs{n},m}
\end{array}\right)
\end{equation}
where
\begin{equation}
\begin{split}
\alpha_{n}=& \begin{cases}
\frac{-i\sign(n)\cos\varphi_{n}}{\sqrt{2(1-\sign(n)\sin\varphi_{n})}}\qquad n\neq 0 \\
\qquad\qquad  0\qquad \qquad n=0
\end{cases}\\
\beta_{n}=&\begin{cases}
\sqrt{(1-\sign(n)\sin\varphi_{n})/2}\qquad n\neq 0 \\
\qquad\qquad  1\qquad \qquad \qquad n=0
\end{cases}
\end{split}
\end{equation}
$\varphi_{n}=g\mu_{B}B/(\hbar v_{F}\ell_{B}^{-1}\sqrt{2\abs{n}})$ and $\sign(n)=n/\abs{n}$. These states have energies~\cite{Mondal2010}
\begin{equation}
\begin{split}
E_{n}=& \sign(n)\sqrt{(g\mu_{B}B)^{2}+2\lvert n\rvert \hbar^{2}v_{F}^{2}\ell_{B}^{-2}} \qquad\text{n $\neq$ 0}\\
E_{0}=& -g\mu_{B}B
\end{split}
\end{equation}
In the limit of $g\rightarrow 0$, the above equations reduce to those of graphene.

The Coulomb interaction is
\begin{equation}
V\left(\lvert {\bm r}_{1}-{\bm r}_{2}\rvert\right)=\frac{e^{2}}{\epsilon\lvert{\bm r}_{1}-{\bm r}_{2}\rvert}
\end{equation}
which acts on the two-body states,
\begin{eqnarray}
\lvert\lvert n\, m_{1}; n\, m_{2}\rangle\rangle &= & \Psi_{n m_{1}}\otimes\Psi_{n m_{2}}\nonumber \\
&=&\alpha_{n}^{2}\lvert \abs{n}-1\, m_{1}\,\up;\abs{n}-1\, m_{2}\, \up\rangle+{}\nonumber \\
&& {}+\alpha_{n}\beta_{n}\lvert \abs{n}\, m_{1}\,\down ;\abs{n}-1\, m_{2}\, \up\rangle+{}\\
&& {}+\alpha_{n}\beta_{n}\lvert \abs{n}-1\, m_{1}\, \up ;\abs{n}\, m_{2}\,\down \rangle+{}\nonumber\\
&& {}+\beta_{n}^{2}\lvert \abs{n}\, m_{1}\, \down ;\abs{n}\, m_{2}\, \down \rangle\nonumber
\end{eqnarray}
where the single ket denotes the product state of the usual quantum well states, $\lvert n_1\, m_1;n_2\, m_2\rangle=\phi_{n_{1}m_{1}}\otimes\phi_{n_{2}m_{2}}$. Then the pseudopotentials of the Coulomb interaction are
\begin{eqnarray}
V_{\rm eff}^{n,m}&=&\langle\langle n\, m_{1};n\, m_{2}\rvert\rvert V\lvert\lvert n\, m_{3};n\, m_{4}\rangle\rangle\nonumber\\
&=&\biggl(\frac{\cos^{4}\varphi_{n}}{4(1-\sign(n)\sin\varphi_{n})^{2}} V_{m}^{(n-1)}+{}\biggr.\nonumber \\
&&\, \biggl.{}+\frac{1}{4}(1-\sign(n)\sin\varphi_{n})^{2}V_{m}^{(n)}+\frac{1}{2}\cos^{2}\varphi_{n} V_{m}^{(n,n-1)}\biggr)\label{Veff}
\end{eqnarray}
where~\cite{Toke2006,JainText}
\begin{eqnarray}
V_{m} ^{(n)}&\equiv &\langle {\abs{n}\, m_1};{\abs{n}\, m_2}\rvert V\lvert {\abs{n}\, m_3}; {\abs{n}\, m_4}\rangle\nonumber  \\
&=&\int\frac{d^{2}k}{(2\pi)^{2}}V(k)\left[L_{\abs{n}}\left(\frac{k^{2}}{2}\right)\right]^{2}L_{m}(k^{2})e^{-k^{2}}\label{Vmn}\\
V_{m}^{(n,n-1)} & \equiv &  \langle {\abs{n}\, m_1};{ \abs{n}-1\, m_2}\rvert V\lvert {\abs{n}\, m_3};{ \abs{n}-1\, m_4}\rangle\nonumber \\
&=&\int \frac{d^{2}k}{(2\pi)^{2}}V(k)L_{\abs{n}}\left(\frac{k^{2}}{2}\right)L_{\abs{n}-1}\left(\frac{k^{2}}{2}\right)L_{m}(k^{2})e^{-k^{2}}\label{Vmmix}
\end{eqnarray}
Here, $V(k)$ is the Fourier transform of the Coulomb potential and $-\hbar m$ is the relative angular momentum.

\begin{figure}
\includegraphics[width=\columnwidth]{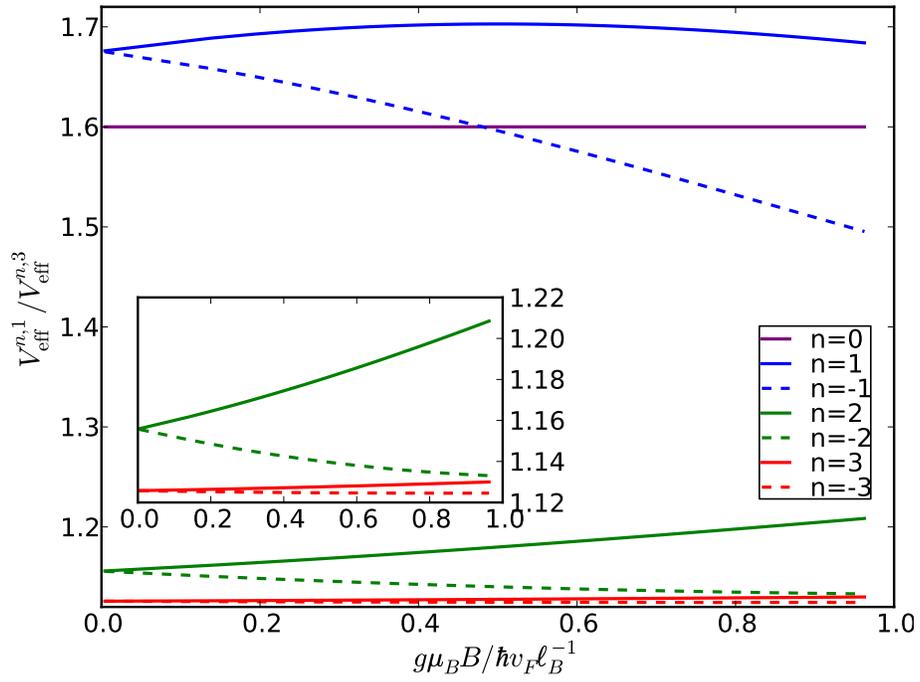}
\caption{The ratio of the first and third relative angular momentum pseudopotentials for the Coulomb interaction in the lowest 7 LLs as a function of Zeeman energy. The solid lines are for positive $n$ while the dashed lines are for negative $n$. The inset is the same as the main figure, but with a different vertical scale to highlight the $\abs{n}=2,3$ LLs.}
\label{ppRatiovsB}
\end{figure}

\section{Results and discussion}

Figure~\ref{ppRatiovsB} shows the ratio of the $m=1$ to the $m=3$ pseudopotentials for the Coulomb interaction as a function of Zeeman energy. In the lowest ($n=0$) LL, the spin is perfectly polarized without the addition of a magnetic field, so the ratio is independent of the Zeeman energy. For all other values of $\abs{n}$, the symmetry between $n$ and $-n$ is broken as the Zeeman energy increases from zero. The pseudopotential ratio increases as a function of Zeeman energy for $n>0$, indicating an increased stability of the fractional quantum Hall state, while just the opposite happens for $n<0$. However, when $n\geq 2$, the pseudopotential ratio becomes smaller than 1.2, and the fractional Hall effect is likely to be unstable. Thus the features seen in Refs.~\cite{Analytis2010,Xiong2011} at filling factors $n>2$ are unlikely to be precursors to the FQHE.

\begin{figure}
\includegraphics[width=\columnwidth]{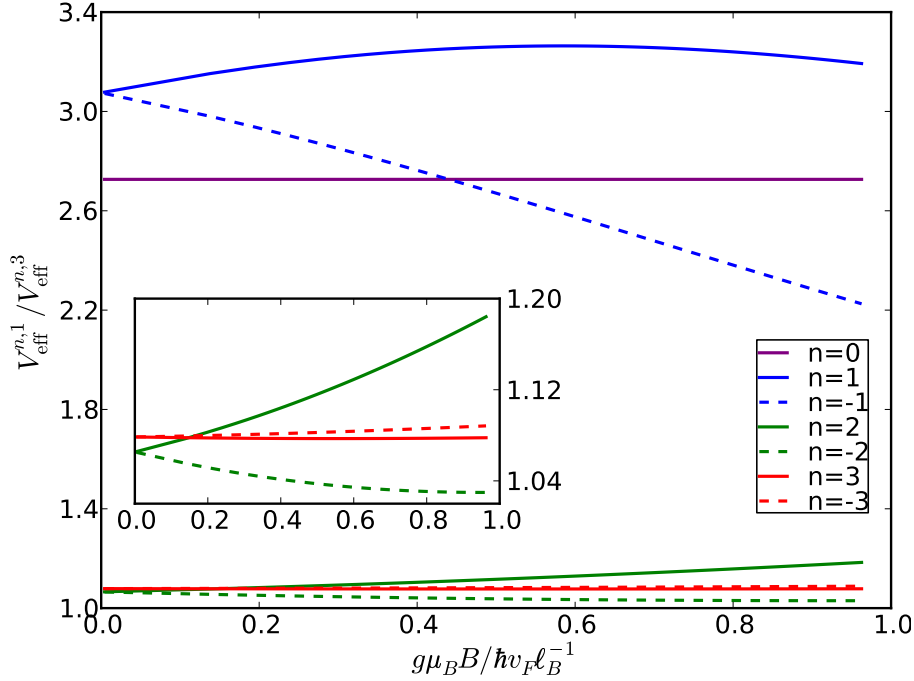}
\caption{The ratio of the first and third relative angular momentum pseudopotentials as a function of Zeeman energy for an interaction which models the effect of screening due to the bulk electrons (see text). The solid lines are for positive $n$ while the dashed lines are for negative $n$. The inset is the same as the main figure, but with a different vertical scale to highlight the $\abs{n}=2,3$ LLs.}
\label{ppRatio-dipole}
\end{figure}

The presence of a noninsulating bulk may significantly alter the form of the interaction. As a model, consider the form
\begin{equation}\label{Vdipole}
V\left({\bm r}\right)=\frac{e^{2}}{\epsilon r \sqrt{1+(r/r_{\rm d})^{4}}}
\end{equation}
which behaves as the Coulomb interaction ($V\sim 1/r$) at short distances and as a dipole-dipole interaction ($V\sim 1/r^{3}$) at large distances. The parameter $r_{\rm d}$ characterizes the size of the dipole. For simplicity, I have taken $r_{\rm d}=1/k_{\rm TF}$; when the two electrons are much further than a Thomas-Fermi screening length, then they interact as if they are two dipoles, with the noninsulating bulk creating the positive end of each dipole. 

The pseudopotential will have exactly the same form as Equations~\eqref{Veff}-\eqref{Vmmix} with $V(k)$ replaced by the Fourier transform of the model interaction,~\eqref{Vdipole}. Figure~\ref{ppRatio-dipole} shows the ratio of the $m=1$ to the $m=3$ pseudopotential for the model interaction. Figures~\ref{ppRatiovsB} and~\ref{ppRatio-dipole} are qualitatively similar, and differ only in quantitative aspects. The ratio, $V_{\rm eff}^{n,1}/V_{\rm eff}^{n,3}$, has increased for the $n=0, 1,$ and $-1$ LLs, which suggests an increased stability of the FQHE. However, the ratio has decreased for $n\geq 2$; i.e. the noninsulating bulk will make the FQHE even less stable.

\section{Conclusion}

In conclusion, this paper has ruled out FQHE in LLs $|n|\geq 2$, while predicting its stability for $n=0,1,-1$ for a TI surface. Due to the strong spin orbit interaction in TIs, there is an asymmetry in the FQHE for $n=1$ and $n=-1$ LLs, in contrast to graphene. The effect of screening by a noninsulating bulk has been included by using a model interaction which looks like a dipole-dipole interaction at distances larger than the Thomas Fermi screening length.  Although the composite fermion states in the LLs $n=0,1,-1$ are further stabilized by this model interaction, the composite fermion state in higher LLs ($\abs{n}\geq 2$) are further destabilized due to the weaker long range interaction.

\section{Acknowledgements}

I would like to thank Jainendra Jain for many useful discussions and for comments on this manuscript. This work was supported by the Penn State MRSEC under NSF grant DMR-0820404.

\end{document}